\documentclass{ptapap}

\usepackage{color}
\usepackage{amssymb}
\usepackage{calrsfs}
\DeclareMathAlphabet{\pazocal}{OMS}{zplm}{m}{n}
\newcommand{\Lb}{\pazocal{L}}

\author{V\'eronique Petit}[UD]
\author{Alexandre David-Uraz}[UD]

\affil[UD]{Dept. of Physics and Astronomy, University of Delaware, Newark, DE, USA}

\title{Recent developments in determining the evolution of magnetic OB stars}

\begin{document}

\maketitle

\begin{abstract}

We review recent developments in determining the evolution of magnetic massive OB stars. One of the important scientific questions is the completeness and the detection limits of contemporaneous spectropolarimetric surveys across the HR diagram. We present the characteristics of the MiMeS survey of O-type stars, and how the limits of the current available observations warrant the design of new high-precision surveys that target the older O-type star population.
Another important question is whether the presence of the magnetic fields changes stellar evolutionary tracks in a significant way, hence leading to a wrong determination of stellar parameters. We review new evolution models that include the effect of magnetic wind quenching, and suppression of convection in the iron opacity peak zone.
\end{abstract}

\section{Introduction}

Let us do a thought experiment. 
A molecular cloud collapses and fragments into a multitude of massive proto-stars. 
In general, simulations of such collapses include the impact of the interstellar magnetic field threading the cloud \citep[e.g.][]{2008MNRAS.385.1820P}. 
It is thus quite natural to expect that in addition to the Initial Mass Function (IMF) that dictates the distribution of stellar masses \citep{1955ApJ...121..161S}, there should also be an Initial B-field Function \citep[I$\vec{\mathrm{B}}$F, ][]{2019MNRAS.489.5669P} that dictates the initial magnetic field strength distribution of newly-born stars. 

Of course, this I$\vec{\mathrm{B}}$F may be a function of parameters that are governed by the exact processes that generate the magnetic fields -- the example above (advection of interstellar magnetic fields)
is only one of the suggested 
hypotheses for the origins of magnetic fields in massive stars. 
Indeed, it has now been established that 1 in 10 OBA stars hosts a large-scale magnetic field that is not contemporaneously generated by a dynamo mechanism similar to that of sun-like stars \citep[e.g. ][]{2007ApJ...657..486B}, but is rather a  
remnant from an earlier event 
or evolutionary phase that has yet to be determined. The most widely accepted hypotheses at the moment involve either a seed magnetic field in the star-forming region as mentioned previously (e.g. \citealt{2001ASPC..248..305M}) or a dynamo acting during the pre-main sequence stage, when these stars go through a fully convective phase \citep{2019A&A...622A..72V}. Whatever scenario one chooses, it must reproduce the apparent lack of correlation between magnetic and stellar properties in massive stars  \citep[e.g.][]{2019MNRAS.490..274S}. 

This said, constraining the I$\vec{\mathrm{B}}$F observationally could provide pointers to the mechanisms that introduce these so-called ``fossil'' fields. Although magnetic fields can now be measured in great detail (see \S 2), reconstituting the I$\vec{\mathrm{B}}$F is however not a straightforward task. We are measuring the current-day characteristics of magnetic massive stars, and being able to rewind the history of the population of known magnetic massive stars requires multiple improvements of our current state of knowledge:
(i) Stellar evolution models have to take the impact of fossil magnetic fields into account, which could for example mean that the age of magnetic stars determined with non-magnetic models may not be accurate. 
(ii) Some field generation scenarios continuously introduce new magnetic fields into a stellar population \citep[e.g. shear-driven dynamos in merger events][]{2019Natur.574..211S} and would have to be properly accounted for when attempting to recover the I$\vec{\mathrm{B}}$F. 
(iii) In order to place the population of magnetic stars in the context of the general population of massive stars, we need to understand the detection limits and the survey biases very well. 

Therefore, the interpretation of current-day magnetic properties of stars to constrain the I$\vec{\mathrm{B}}$F is entangled with the uncertain magnetic stellar evolution, and the uncertain observational biases.

\section{Have We Observed All Possible Groups of Massive Stars and Have We Observed Them Well Enough?}

The Zeeman effect \citep{1897ApJ.....5..332Z} provides two ways of detecting magnetic fields in massive stars -- the Zeeman splitting of spectral lines into multiple components, and the difference in polarization state across those components (usually in circular polarization, thus the Stokes V parameter).  
In massive stars, the Zeeman splitting is usually too small to be detected against other sources of line broadening (e.g. rotation, macro-turbulence). 
Therefore spectropolarimetry, combined with techniques that consider 
the Stokes V signal in multiple spectral lines at once \citep{1997MNRAS.291..658D}, is the modern way to detect magnetic fields. For a review, see \citet{2009ARA&A..47..333D}. 

\begin{figure}
\includegraphics[width=\textwidth]{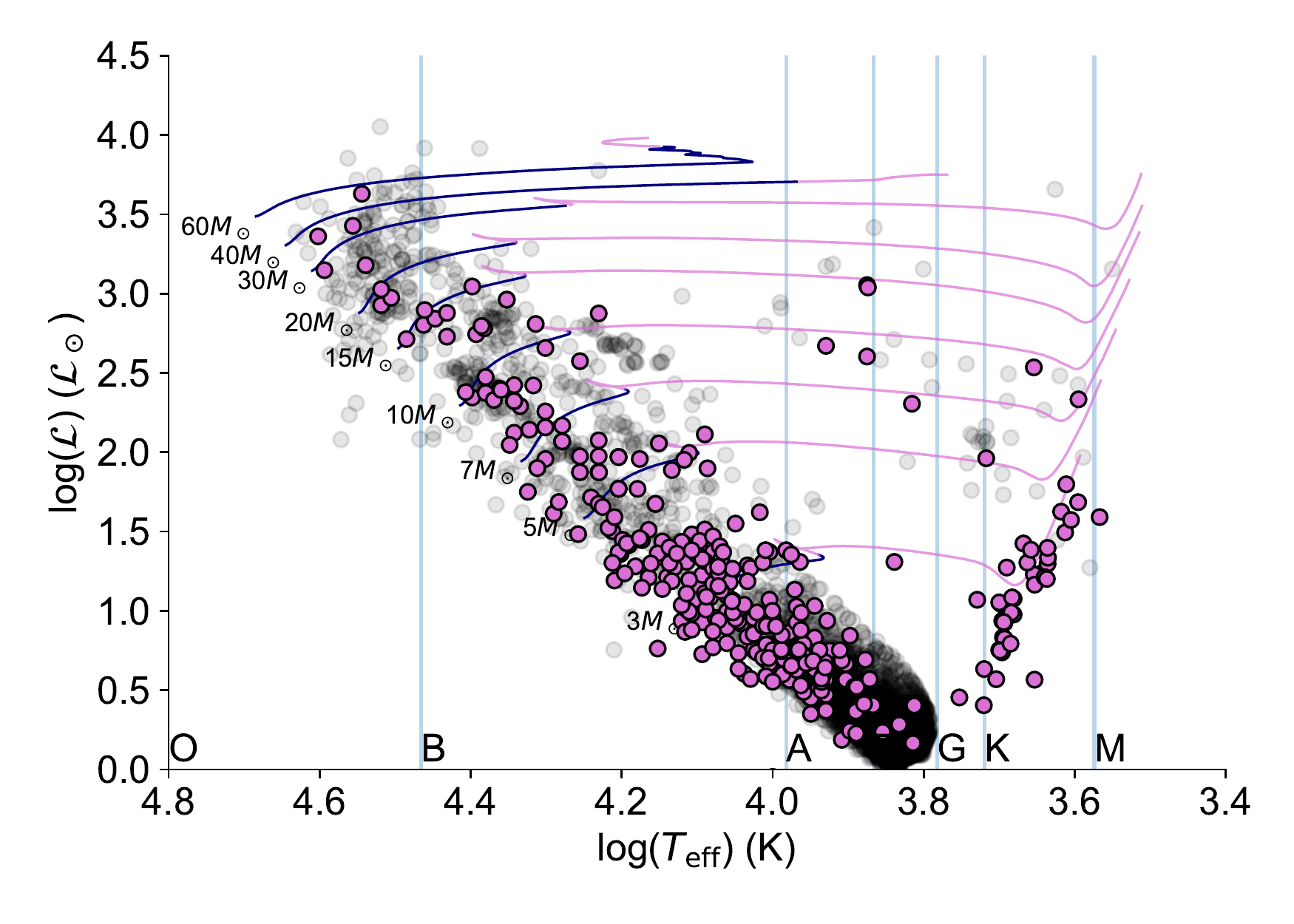}
\caption{\label{fig:hrd}Spectroscopic HRD  
for detected (pink) and non-detected (black/grey) stars in a collection of spectropolarimetric surveys (see text for details). For the non-detections, the data points have an opacity of 0.1 to illustrate density of coverage. The evolution tracks are from the non-rotating models of \citet{2011A&A...530A.115B}, and are colored such that the main sequence is in dark blue and the post-main sequence in light pink. The approximate spectral types are shown with vertical lines.  }
\end{figure}

In this section, we provide a non-exhaustive overview of the regions in the Hertzsprung-Russell diagram (HRD)
that have, to date, been covered with large spectropolarimetric surveys. Figure \ref{fig:hrd} shows the upper left region of the so-called spectroscopic HRD
\citep[sHRD;][]{2014A&A...564A..52L} that displays $\log{\Lb} \equiv$ $\log( T_\mathrm{eff}^4/g)$ versus $\log T_\mathrm{eff}$. We show a set of non-rotating evolution tracks from \citet{2011A&A...530A.115B} as a reference. 

The datapoints are taken from multiple surveys of OBA stars, with magenta color representing magnetic stars and grey color representing stars that were observed with spectropolarimetry but for which no Stokes V signal was found. We note here that we have not made any attempt 
to assess the upper limits of these non-detections -- therefore this figure must be taken as an illustration of survey selection bias, but should not be used to assess the incidence of magnetism.  The surveys used are as follow:
\begin{itemize}
\item All the ApBp stars from the cluster surveys of \citet{2006A&A...450..777B} and \citet{2007A&A...470..685L,2008A&A...481..465L}. As the surveys concentrated on known chemically peculiar stars, this dataset only contains magnetic detections. 

\item The volume-limited sample of \citet{2019MNRAS.483.2300S,2019MNRAS.483.3127S} that contains all the intermediate-mass main sequence (MS) stars within 100pc. 

\item The red giant sample of \citet{2015A&A...574A..90A}, which likely includes dynamo-generated fields but also descendants of ApBp stars. 

\item The sample of late-type supergiants from \citet{2010MNRAS.408.2290G}. 

\item The evolved magnetic massive stars reported by \citet{2017MNRAS.471.1926N} and \citet{2018MNRAS.475.1521M} in the context of the Large Impact of magnetic Fields on the Evolution of hot stars (LIFE) project. 

\item All the currently known magnetic OB stars from the compilations of \citet{2013MNRAS.429..398P} and \citet{2016A&A...592A..84F}. 

\item All the O and B stars observed in the context of the MiMeS survey \citep{2016MNRAS.456....2W,2017MNRAS.465.2432G,2019MNRAS.489.5669P}.

\end{itemize}

Three other significant surveys were not included here -- the BOB survey of $\sim$100 MS B-type stars \citep{2017A&A...599A..66S}, the compilation of FORS1 data of \citet{2015A&A...583A.115B}, and the BinaMIcS survey of $\sim$150 binary systems \citep{2015IAUS..307..330A} -- because a tabulation of the spectral stellar parameters 
was not readily available. 

As we can see in Fig.~\ref{fig:hrd}, most observed magnetic stars are still on the main sequence, and there is especially a lack of evolved magnetic stars with higher masses ($\gtrsim 15$~M$_\odot$). This observation could be caused by a combination of factors.
First, there have been less large-scale efforts to observe very evolved stars and that region of the HRD is less covered by spectropolarimetric surveys, leading to survey selection biases that are not well characterized.

Secondly, while magnetic detections are published regularly, there have been few  
studies characterizing the detection limits across the HRD based on actual survey data. This is especially important for massive OB stars, as the pre-selection of magnetic candidates cannot entirely rely on
spectral peculiarities like it is the case for ApBp stars. For example, magnetic B-type stars can be either helium strong or weak  
\citep{1979ApJ...228..809B,1983ApJS...53..151B}, or show no helium 
peculiarity \citep{2006MNRAS.370..629D,2011MNRAS.412L..45P}. Furthermore, while the Of?p spectral type classification has been directly associated with magnetism \citep{2010ApJ...711L.143W}, 
not all magnetic O-type stars are Of?p \citep{2009MNRAS.400L..94G}. 

\begin{figure}
\includegraphics[width=\textwidth]{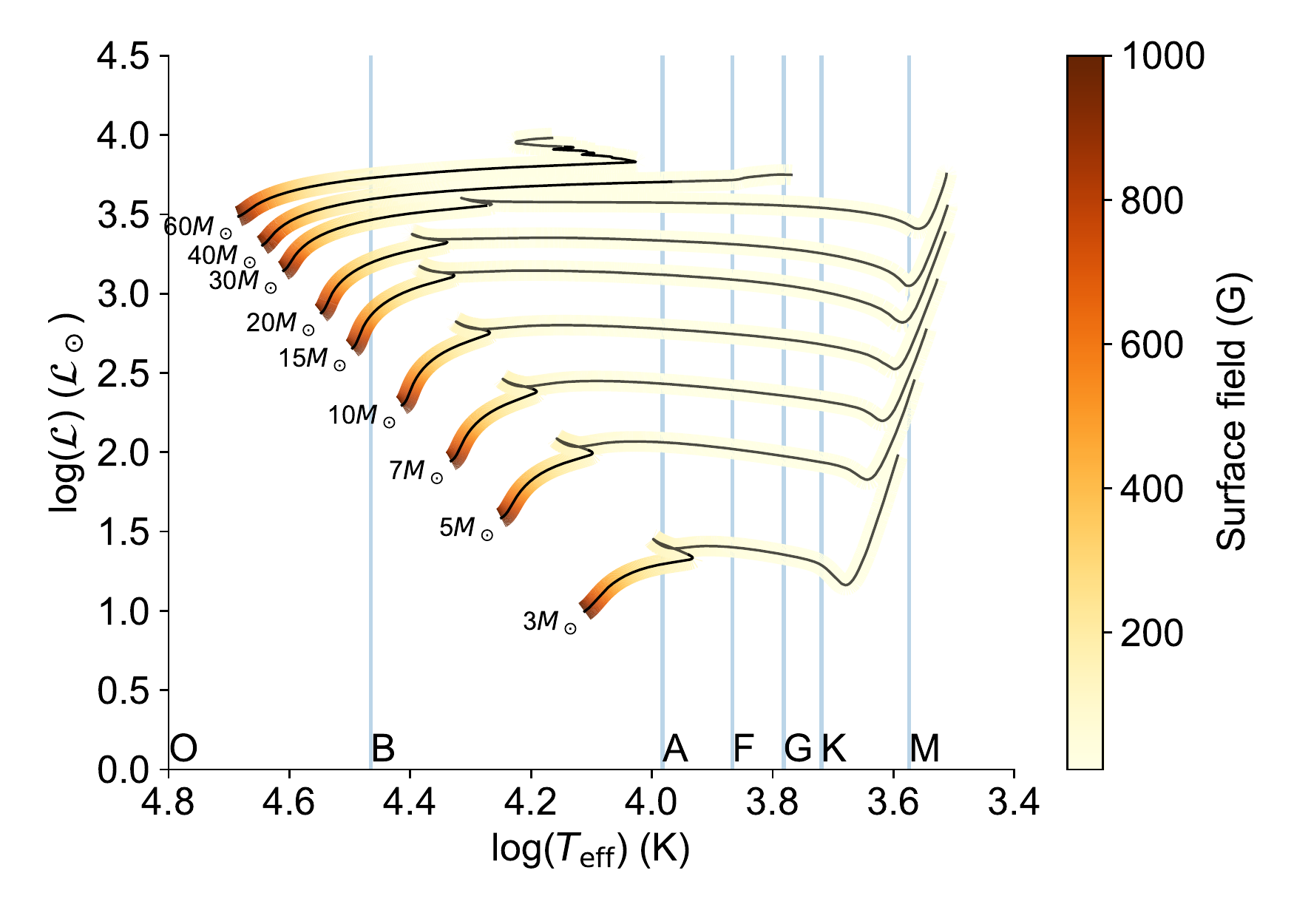}
\caption{\label{fig:detect} Evolution tracks \citep{2011A&A...530A.115B} in the sHRD, color-coded according to the expected surface field strength under the hypothesis of magnetic flux conservation, with the ZAMS field set to 1~kG. We note that these models do not include the feedback of the fossil field on the stellar structure and evolution. Nevertheless, it illustrates the large drop in magnetic field strength as the star evolves off the main sequence. The approximate spectral types are shown with vertical lines.  }
\end{figure}

Finally, the magnetic fields of evolved massive stars are likely inherently harder to detect. Under the magnetic flux conservation assumption, the surface field strength is expected to be inversely proportional to the square of the stellar radius. Thus as a star ages and its radius increases, we can expect the magnetic strength to decrease rapidly. Figure~\ref{fig:detect} illustrates the decrease in surface field expected for stars of various masses with a 1~kG magnetic field at the ZAMS (note that these tracks do not include evolutionary feedback from the presence of the magnetic field, see next section). As can be seen, the detection threshold is considerably lower at the TAMS than at the ZAMS, and while the fields on many of these stars might not be easily detected, a careful analysis of the upper limits of non-detections can constrain the magnetic properties of the population.

The MiMeS Survey \citep{2016MNRAS.456....2W} has performed such an analysis using the upper limits achieved for its sample of O-type stars \citep{2017MNRAS.465.2432G, 2019MNRAS.489.5669P} to determine the range of dipolar field strengths permitted by the polarisation spectra. They established that a model in which all the stars in the sample were to host a 100~G, dipolar magnetic field can be ruled out by the MiMeS data. Of particular relevance here, they also find that better upper limits, by at least a factor of 10, would have been necessary to rule out a detection bias as an explanation for the apparent lack of evolved main-sequence magnetic O-type stars \citep{2016A&A...592A..84F}. 

Therefore dedicated surveys, such as the LIFE survey \citep{2018MNRAS.475.1521M}, targeting the end of the main sequence for O-type stars with very high magnetic precision are necessary to confirm or refute the flux conservation hypothesis for massive stars.

\section{Evolution of Magnetic Massive Stars}

Another important aspect in determining the evolution of magnetic massive stars is the inclusion of the impact of a fossil field onto stellar structure in evolution models. The effect of fossil magnetism can be separated into two main categories: internal effects and wind-field interaction. 
The magnetic fields in stellar interiors cannot be observed directly. Nevertheless, we can rely on the theoretical description of the type of magnetic field topologies that can exist in the interiors of massive stars \citep[e.g.][]{2010A&A...517A..58D}. Furthermore, asteroseismology of pulsating magnetic B-stars provides indirect evidence that these stars have lower core overshooting than similar non-magnetic stars \citep{2012MNRAS.427..483B}. This could indicate that fossil fields penetrate deep into the interiors, and even interact with the dynamo fields that exist in the cores \citep{2009ApJ...705.1000F,2016ApJ...829...92A}. Another example of indirect evidence is the magnetic suppression of macro-turbulent broadening in hot stars \citep{2013MNRAS.433.2497S,2019MNRAS.487.3904M}.  

The other impact of fossil magnetism is the interaction with the stellar winds. If the field has a large-scale component extending well above the stellar surface, it channels the wind material, creating a structured circumstellar magnetosphere that (i) quenches the total mass-loss, as the trapped material falls back to the star unless centrifugal support is present \citep{2008MNRAS.385...97U} and (ii) increases the angular momentum loss by magnetically braking the surface
\citep{2009MNRAS.392.1022U}. \citet{2011A&A...525L..11M} considered the effect of a fossil magnetic field on angular momentum loss in the evolution of a 10~M$_\odot$ star. Considering two models reflecting extreme behaviors of angular momentum transport in the interior, they found that when the interior is in solid-body rotation, the surface rotation decreases more rapidly with time than when the interior is differentially rotating.

More recently, \citet{2017MNRAS.466.1052P} implemented the quenching of the mass loss and the time evolution of the field based on the magnetic flux conservation hypothesis.  They found that for stars above 40~M$_\odot$, magnetic quenching of stellar winds significantly modifies stellar evolutionary tracks even during the MS. For example, a 80~M$_\odot$ star with a field strength like that of NGC 1624-2 \citep[the most magnetic O-type star known to date;][]{2012MNRAS.425.1278W} would reach the TAMS with a significantly larger mass (by $\sim$20~M$_\odot$) than 
a non-magnetic star of the same initial mass. This could present another pathway for the creation of “heavy” stellar-mass black holes such as those whose coalescence was detected by LIGO \citep{2016PhRvL.116f1102A}, even at solar metallicity.
Another interesting application of magnetic quenching is to facilitate the formation of pair-instability supernovae at solar metallicity \citep{2017A&A...599L...5G}.

\citet{2019MNRAS.485.5843K} implemented the effect of wind quenching, magnetic braking, and field strength evolution in lower initial mass models, for which wind mass-loss is less significant for the mass evolution of the stars. Nevertheless, the effect of magnetic braking was found to be significant. For two extreme schemes for the transport of angular momentum in the interior (uniform removal of angular momentum throughout the radiative envelope and removal only from the surface layers) the magnetic star’s surface rotation slows down significantly in only a few Myr.

\section{Summary}

We presented what we hope to be a compelling case for empirically determining the Initial B-field Function (I$\vec{\mathrm{B}}$F) of stars with a radiative envelope. We presented the latest developments with respect to various problems that hamper the determination of the I$\vec{\mathrm{B}}$F. First, the observational biases and detection limits of large spectropolatimetric surveys remain to date poorly constrained. On the bright side, there have
been new theoretical improvements in modelling the evolution of massive magnetic stars.

We will soon be able to use population studies, combining measured magnetic characteristics (corrected for biases) with appropriate evolutionary models to constrain the I$\vec{\mathrm{B}}$F \citep[e.g.][these proceedings]{2020arXiv200110564C}.

\acknowledgements{
The authors would like to thank John Landstreet, Gregg Wade, Luca Fossati, and James Sikora for providing the tabulated data included in Fig.~\ref{fig:hrd}. 
VP acknowledges support from the University of Delaware Research Foundation. This material is based upon work supported by
the National Science Foundation under Grant No.~1747658. ADU acknowledges support from the Natural Sciences and Engineering Research Council of Canada (NSERC).}

\bibliographystyle{ptapap}
\bibliography{database}

\end{document}